\documentclass[multphys,vecphys]{svmult}

\usepackage{makeidx}         % allows index generation
\usepackage{graphicx}        % standard LaTeX graphics tool
\usepackage{multicol}        % used for the two-column index
\usepackage[bottom]{footmisc}% places footnotes at page bottom
\usepackage{natbib}

\makeindex             % used for the subject index
\begin{document}

\title*{The Primordial Binary Population in OB~Associations}
\author{Thijs Kouwenhoven}
\institute{Astronomical Institute `Anton Pannekoek',
   University of Amsterdam,
   Kruislaan 403, 1098 SJ Amsterdam, The Netherlands
\texttt{kouwenho@science.uva.nl}
}
%\titlerunning{xxxx}
%\authorrunning{xxxxx}
\maketitle

\begin{abstract}
  We describe our method to find the primordial binary population in the Sco~OB2 association. We present the results of our ADONIS and VLT/NACO near-infrared adaptive optics binarity surveys of A and late-B stars in Sco~OB2. We combine these results with literature data on visual, spectroscopic, and astrometric binary stars. With our observations we remove part of the selection effects present in the combined dataset. Using simulated observations the remaining biases can be removed in order to derive the true binary population. Detailed N-body simulations, including stellar and binary evolution are required to derive constraints on the binary population which was present just after the natal gas has been removed from Sco~OB2, i.e., the primordial binary population.  
\end{abstract}

\section{Introduction}
\label{sec:1}
% Always give a unique label
% and use \ref{<label>} for cross-references
% and \cite{<label>} for bibliographic references
% use \sectionmark{}
% to alter or adjust the section heading in the running head

Over the last decade observations of star forming
regions and young stellar clusters have shown that most,
possibly even all stars form in binaries or higher-order multiple
systems. For our understanding of the star
and cluster formation process it is important to characterize
the properties of binary stars as a function of, e.g., the mass of the primary star and
environment. Ideally, one would like to know all properties of binary and multiple systems just after the stars have formed. Unfortunately, performing a binarity study in star forming regions poses severe observational difficulties due to the large amount of interstellar gas and dust. However, shortly after the first massive stars are formed, their stellar winds rapidly remove the gas and dust from the star-forming region. From this point in time, accretion ceases almost completely and the binary stars do not interact with the gas anymore. After the gas has been removed the dynamical evolution of the binary population is expected to become much less important.

The primordial binary population (PBP) is {\em the population of binaries as established just after the gas has been removed from the forming system, i.e., when the stars can no longer accrete gas from their surroundings} \citep{kouwenhoven_kouwenhoven2005}. 
The evolution of stars and dynamics of the newly born stellar population is influenced by the presence of gas, but when the gas is removed, the binary population is only affected by stellar evolution and pure N-body dynamics. The PBP is also a natural boundary for simulations of the formation of star clusters and binary stars. Hydrodynamical simulations of a contracting gas cloud \citep[e.g.,][]{kouwenhoven_bate2003} result in the primordial binary population when the gas is removed by accretion, stellar winds, or supernova explosions. Pure N-body simulations \citep[e.g.,][]{kouwenhoven_ecology4} can be used to study the subsequent evolution of the star cluster and binary population.

OB~associations are well suited for studying the PBP. They are young ($5-50$~Myr) and low-density ($< 0.1 M_\odot {\rm pc}^{-3}$) aggregates of stars. Their youth implies that only a handful of the most massive systems have changed due to stellar evolution. The effects of dynamical evolution are expected to be limited due to the young age and low stellar density of the association. Moreover, in contrast to, e.g., the Taurus T~association, OB associations cover the full range of stellar masses.

Scorpius~OB2 is the closest young OB~association and the prime candidate for studying the binary population. The proximity of Sco~OB2 ($118-145$~pc) facilitates observations. The young age ($5-20$~Myr) suggests that dynamical evolution has not significantly altered the binary population since the moment of gas removal. Sco~OB2 consists of three subgroups: Upper Scorpius (US), Upper Centaurus Lupus (UCL), and Lower Centaurus Crux (LCC). The three subgroups are located at a distance of 145~pc, 140~pc, and 118~pc, and have an age of 5~Myr, $15-22$~Myr, and $17-23$~Myr, respectively. The membership and stellar content of the association was established by \cite{kouwenhoven_dezeeuw1999} using {\em Hipparcos} parallaxes and proper motions.
The structure of the Sco~OB2 complex is likely the result of sequential star formation. The LCC and UCL subgroups are the oldest, and may have triggered star formation in US, which in turn may have triggered star formation in the star forming region $\rho$~Oph \citep{kouwenhoven_degeus1992,kouwenhoven_preibisch1999}. By studying the properties of the binary population in the three different subgroups in Sco~OB2, one can establish whether the binary population has evolved as a function of time. 

Our ultimate goal is to derive the PBP in Sco~OB2. Our strategy is to (i) collect data on binarity in Sco~OB2 using observations and literature data, (ii) correct for the selection effects introduced by the different observing techniques, and (iii) use inverse dynamical population synthesis \citep[e.g.,][]{kouwenhoven_kroupa1995} to derive the PBP.

\section{The Observed Binary Population}

\begin{figure}[bt]
  \begin{tabular}{cc}
    \includegraphics[width=0.5\textwidth,height=!]{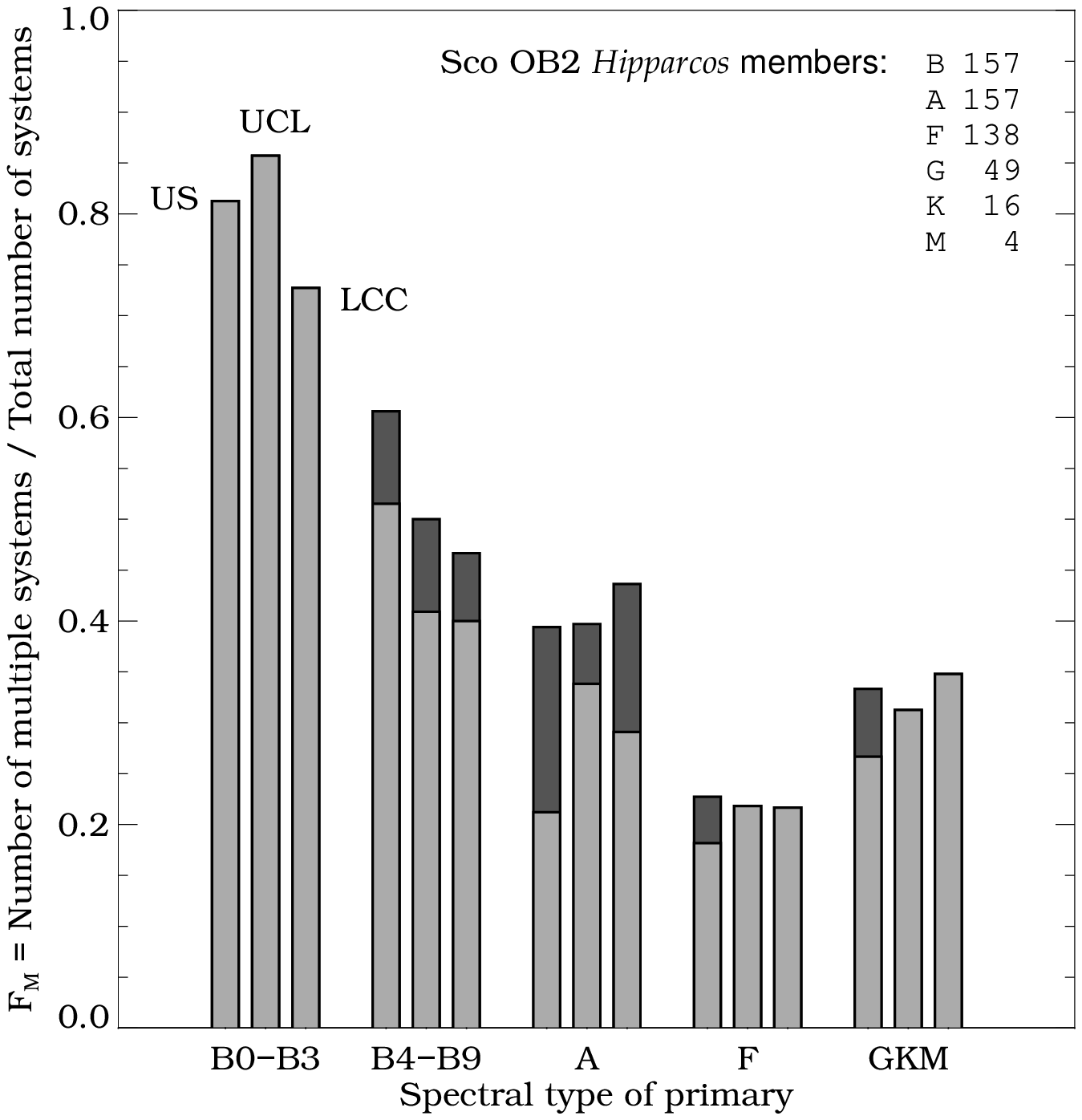} &
    \includegraphics[width=0.5\textwidth,height=!]{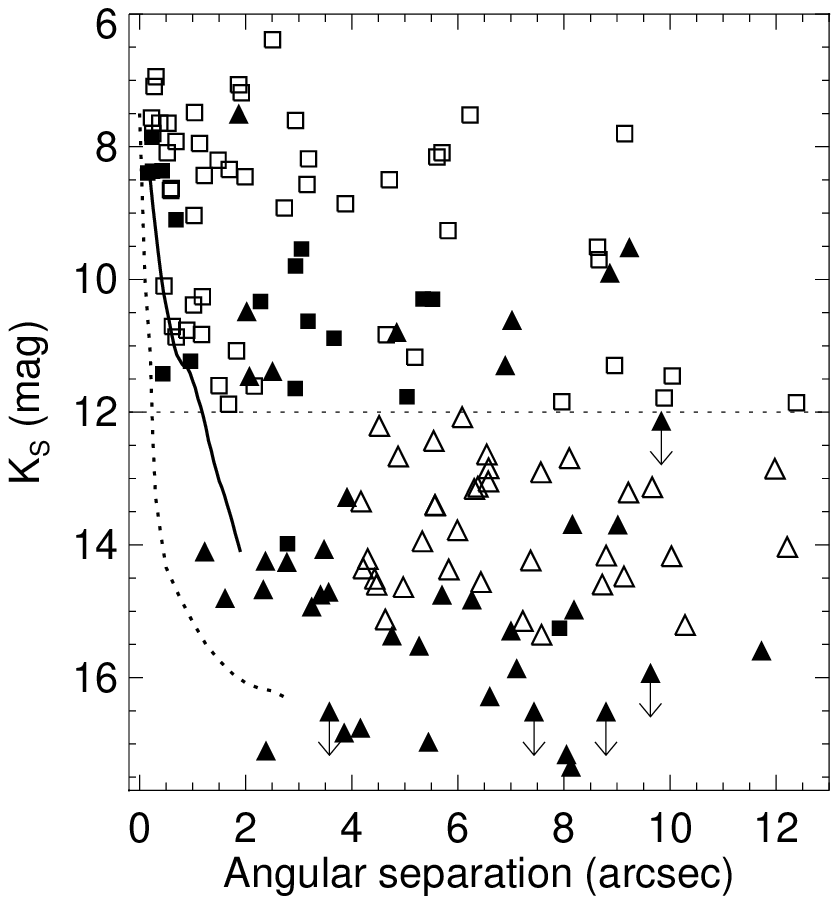} \\
  \end{tabular}
  \caption{{\em Left:} The fraction of stellar systems which is multiple versus the
  spectral type of the primary, for the three subgroups of Sco~OB2. 
  The light and dark
  gray parts of the bars correspond to literature data and the new data
  presented, respectively. The spectral types of the companion
  stars (not included in this plot) are always later than those of the primary
  stars. Apparently, the multiplicity is a function of spectral type, but this
  conclusion may well be premature when observational biases are not properly
  taken into account. 
  {\em Right:} The results of our ADONIS and VLT/NACO binarity surveys amongst A and late-B members of Sco~OB2, showing the confirmed and candidate companions (closed and open squares), as well as the confirmed and candidate background stars (closed and open triangles). The solid and dotted curves are estimates for the detection limit of the ADONIS and NACO surveys, respectively. Only one brown dwarf companion ($K_S > 12$~mag) is found between $1''$ and $4''$ in the sample of 199 Sco~OB2 members, which provides supporting evidence for the existence of a brown dwarf desert for A and late-B stars in Sco~OB2.
  }
\end{figure}

Brown~(2001) performed an extensive literature search on binarity in Sco~OB2, including visual, spectroscopic, and astrometric binaries. A drastic decline is seen in binary fraction going from early-type to late-type stars stars, which may very well be due to observational selection effects (Figure~1). This selection effect has at least partially been removed by the B-star adaptive-optics binarity survey of \cite{kouwenhoven_shatsky2002}. Anticipating on to finding many new companion stars, we performed a near-infrared adaptive optics survey amongst A and late-B stars in Sco~OB2. Near-infrared observations are preferred over optical observations because of the much smaller luminosity contrast between a massive primary and a low-mass companion. With adaptive optics observations we bridge the gap between the known close spectroscopic binaries and wide visual binaries.

We carried out a near-infrared adaptive optics survey among 199~A and late-B members of the Sco~OB2 association \citep{kouwenhoven_kouwenhoven2005}. In total 151 secondaries are detected, with angular separation $0.22'' \leq \rho \leq 12.4''$ and $6.4 \leq K_S \leq 15.4$~mag. Using a brightness criterion the secondaries are separated into 77 probable background stars and 74 candidate companions (of which 41 previously undocumented). We use 5~Myr and 20~Myr (for US and UCL/LCC, respectively)  isochrones to derive masses and mass ratios. The mass ratio distribution follows $f(q) \propto q^{-0.33}$. Random pairing between primaries and companions is excluded (even if we would assume the background stars to be companions) in the observed primary mass and angular separation regime. No close ($\rho \leq 3.75''$) companions are found in the magnitude range $12 < K_S < 14$~mag, but several close probable background stars with $K_S > 14$~mag are found. The non-detection of close companions with $12 \leq K_S \leq 14$~mag indicates the absence of close brown dwarf companions. 

Seven close ($\rho < 3.75''$) candidate background stars with $K_S > 14$~mag were detected in the ADONIS survey. In order to find the nature of these objects, and to confirm the status of several doubtful candidate companions, we performed VLT/NACO $JHK_S$ follow-up observations. The sample consists of 22~members of Sco~OB2, including 7 next to which the close candidate background stars were found, the others with candidate companions. Our observations were sensitive to companions with $0.1'' < \rho < 12''$ and $K_S < 17$~mag. The multi-color observations allow us to place the secondaries in the color-magnitude diagram, and to compare their position to the isochrone. The seven close candidate background stars are indeed background stars. Six doubtful candidate companions found in the ADONIS survey turn out to be background stars. We confirm the status of the 18 candidate companions and 44 background stars. We find two brown dwarf companions of HIP81972, at an angular separation of $7.92''$ (1500~AU) and $2.97''$ (520 AU)  with a mass of $32~M_J$ and $63~M_J$, respectively. The $63~M_J$ companion is the only brown dwarf detected between $1'' \leq \rho \leq 4''$ (Figure~1). One cannot make a similar statement outside this angular separation range since brown dwarfs with $\rho < 1''$ are undetected in the wings of the primary PSF, and for $\rho > 4''$ the status of many background stars is only statistically confirmed. For a semi-major axis distribution of the form $f(a) \propto a^{-1}$, about 12\% of the companions (thus also 12\% of the brown dwarf companions, assuming that companion mass and $a$ are indepedent) is expected to have $1'' \leq \rho \leq 4''$. 

In our survey among 199~A and late-B stars in Sco~OB2 we find only {\em one} brown dwarf companion in the angular separation range $1''-4''$, although we should have detected brown dwarfs around all stars in this range, if present. This implies a virtual absence of $0.007~M_\odot < M < 0.1~M_\odot$ ($7~M_J < M < 105 M_J$) companions of A and late-B stars with semi-major axes of $\sim 150-550$~AU. The brown dwarf companion fraction for $\sim 150-550$~AU is $0.5 \pm 0.5\%$, which is much smaller than the stellar companion fraction of $14 \pm 3\%$ in this range. {\em Our results provide supporting evidence for the existance of a brown dwarf desert for~A and late-B stars}. 

We combine the results of our ADONIS and NACO surveys with all available literature data on binarity in Sco~OB2, including visual, spectroscopic, and astrometric binaries. The {\em observed} binary fraction shows a clear correlation with the spectral type of the primary (Figure~1), but our results indicate that this trend is at least partially due to selection effects. The observed companion star fraction $F_M \equiv (B+2T+\dots)/(S+B+T+\dots)$ slightly decreases with the age of the subgroups, with a value of 0.61 for US, 0.52 for UCL, and 0.45 for LCC. The overall observed companion star fraction of Sco~OB2 is 0.52.

\section{The True Binary Population}

\begin{figure}[bt]
  \begin{tabular}{cc}
    \includegraphics[width=0.5\textwidth,height=!]{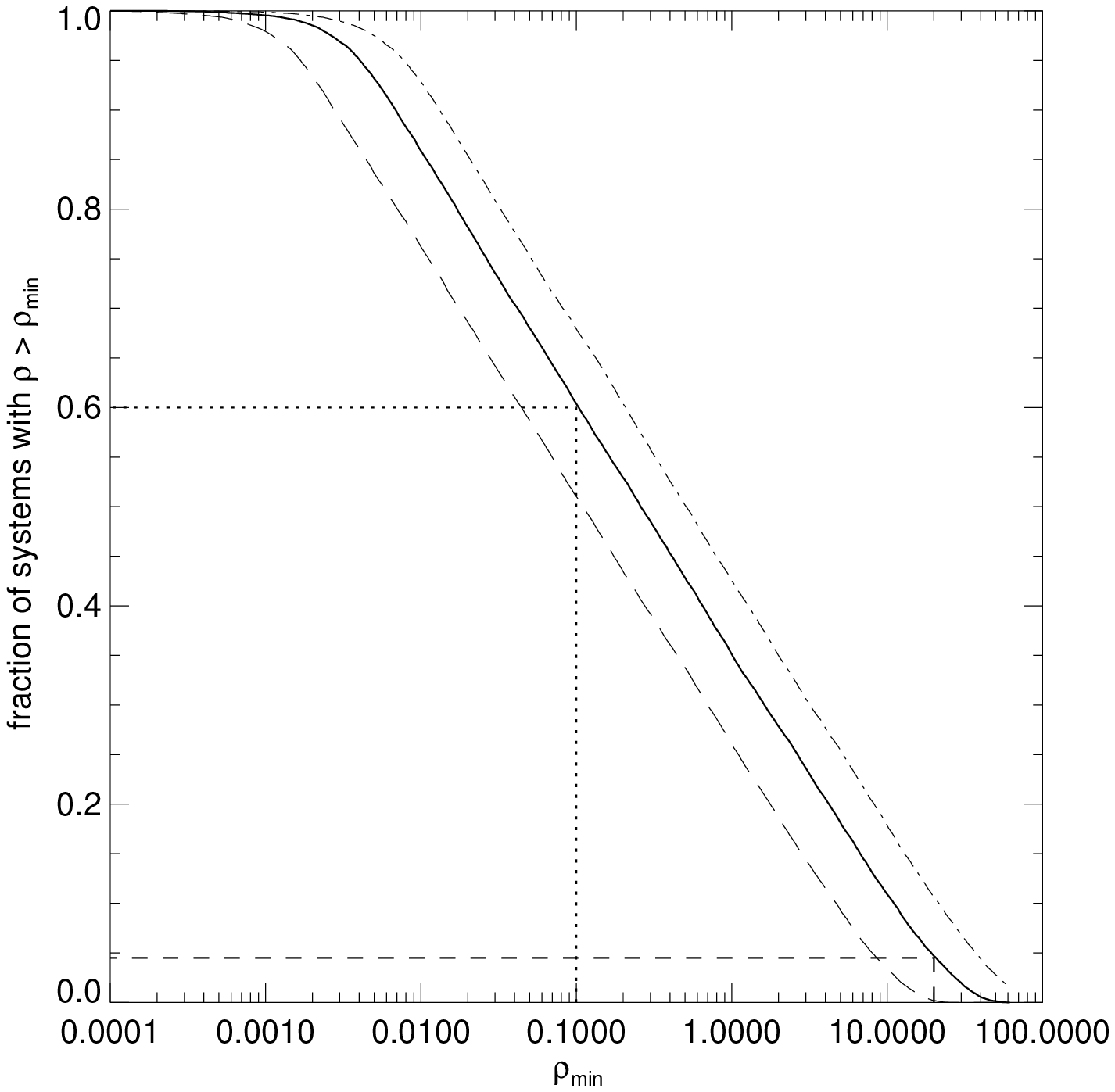} &
    \includegraphics[width=0.5\textwidth,height=!]{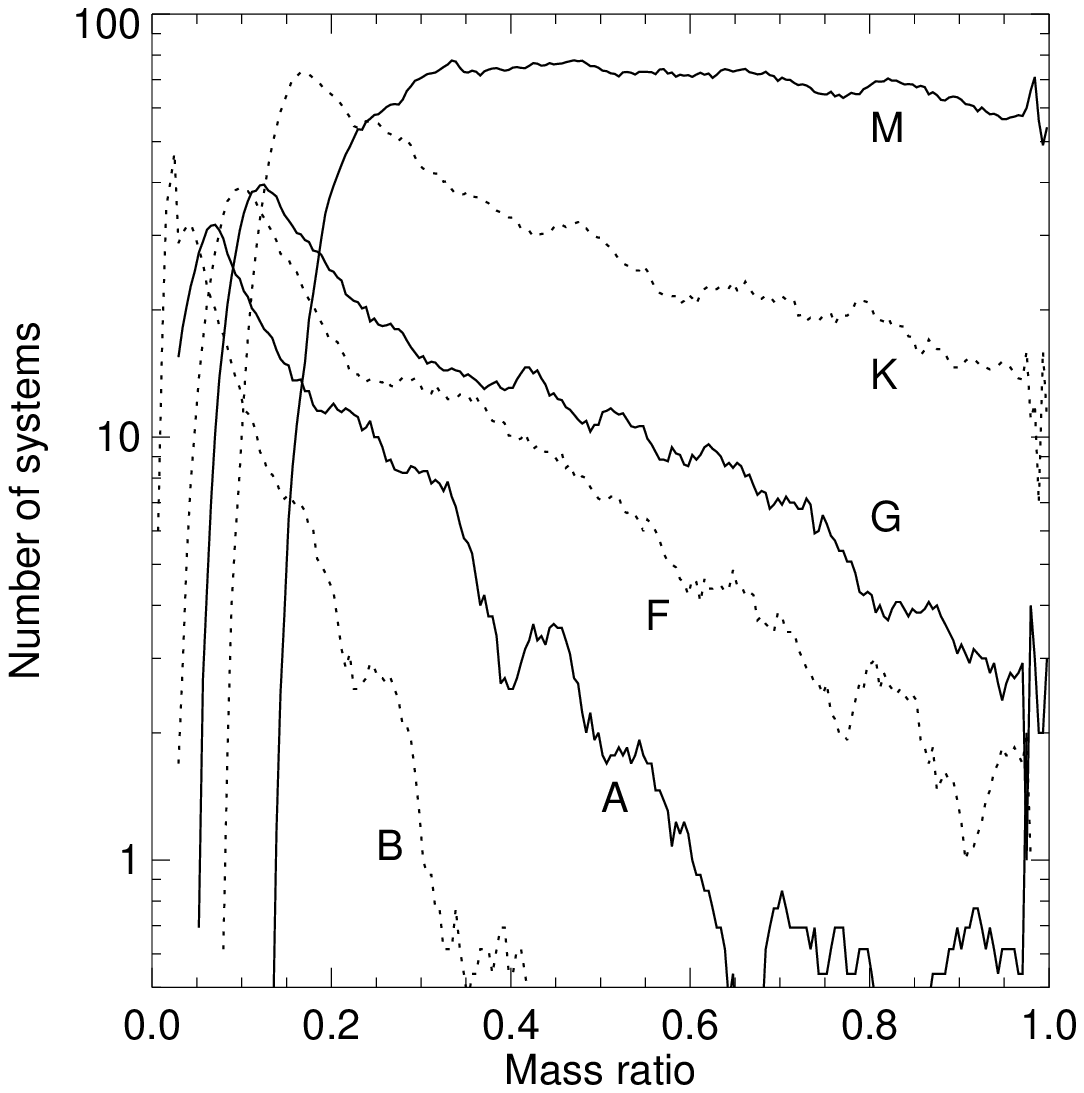} \\
  \end{tabular}
  \caption{{\em Left:} The fraction of binary systems with angular separation $\rho > \rho_{\rm min}$ as a function of $\rho_{\rm min}$. An association with $25,000$ binaries at a distance of 145~pc (solid curve) is used. The association has a thermal eccentricity distribution and a semi-major axis distribution of the form $f(\log a) = \mbox{constant}$ between $10^2~R_\odot$ and  $10^6~R_\odot$, where the limits correspond to the distances at which Roche lobe overflow and Galactic tidal forces become important.  
    In our ADONIS and NACO surveys we can measure angular separations between $\sim 0.1''$ (dotted line) and $\sim 20''$ (dashed line), and hence about 55\% of the companion stars. 
    The observable fraction of binary systems varies slowly with distance, as is illustrated for the models with a distance much closer (75~pc, dash-dotted curve) or farther (350~pc, long-dashed curve) than that of Sco~OB2 .
    {\em Right:} The mass ratio distribution resulting from random pairing between primary and companion star from the Preibisch mass function, split up by primary spectral type. The mass range of the stars is $0.08~M_\odot \leq M \leq 20~M_\odot$. This figure clearly illustrates the danger of comparing mass ratio distributions resulting from binarity surveys of different primary spectral types. 
  }
\end{figure}

Our binarity dataset contains measurements obtained with a wide range of techniques and instruments, each with their specific observational biases. In order to interpret Figure~1 correctly, and to get the {\em true} (i.e., not {\em observed}) binary parameters in Sco~OB2 (or any other population of binaries), it is of crucial importance to understand all selection effects. 
It is impossible to observe the complete binary parameter space in Sco~OB2 due to observational and time constraints, but it is possible to estimate the true binary population using simulations. We use sophisticated simulation techniques to characterize the selection effects and to find the true binary population of Sco~OB2.

The most prominent observational selection effects are the constraints on angular separation $\rho$. With our surveys we are sensitive to $\sim 0.1'' < \rho \sim 20''$, and can only detect companions within this range. Even though these constraints seem rather strict, we are still able to detect $50-55\%$ of the companion stars in our surveys (Figure~2). 

Another important selection effect results from the properties of the surveyed sample. For example, the observed mass ratio distribution depends strongly on the spectral type of the primaries on the sample, even when other selection effects are ignored.
The same intrinsic pairing function (e.g., random pairing) can result in different mass ratio distributions for surveys of stars with different spectral types (Figure~2). For the comparison between mass ratio distributions resulting from different samples (e.g., binarity survey among solar-type stars by \cite{kouwenhoven_duquennoy1991} and our early-type binarity survey), one has to carefully study the sample selection effect.

We will investigate the selection effects resulting from observational constraints, sample selection, and background star contamination using detailed simulated observations. We will perform this analysis for visual, spectroscopic, and astrometric binarity surveys, which will give us the true binary population.

The next step will be to derive the primordial binary population from the true binary population. We will use the STARLAB package \citep{kouwenhoven_ecology4}, which uses state-of-the-art N-body simulations, stellar evolution, and binary evolution. Using inverse dynamical population synthesis \cite[e.g.,][]{kouwenhoven_kroupa1995} we will find the primordial binary population that corresponds most to the true binary population.

\section*{Acknowledgments}

The author wishes to thank Anthony Brown, Lex Kaper, Simon Portegies Zwart, and Hans Zinnecker for their crucial roles in making this project possible. The author also thanks the Leids Kerkhoven Bosscha Fonds for providing financial support for participation in this workshop.

%
% BibTeX users please use
 \bibliographystyle{aa}
 \bibliography{kouwenhoven_bibliography.bib}
%
% Non-BibTeX users please follow the syntax
% the syntax of "referenc.tex" for your own citations
%\input{referenc}
%%%%%%%%%%%%%%%%%%%%%%%%%%%%%%%%%%%%%%%%%%%%%%%%%%%%%%%%%%%%%%%%%%%%%%  }

%%%%%%%%%%%%%%%%%%%%%%%%%%%%%%%%%%%%%%%%%%%%%%%%%%%%%%%%%%%%%%%%%%%%%%

\printindex
\end{document}